\documentclass[runningheads]{llncs}

\usepackage{amstext}
\usepackage{times}
\usepackage{helvet}
\usepackage{courier}
\usepackage{amsmath, amssymb}
\usepackage{latexsym}
\usepackage{eurosym}
\usepackage{wasysym}

\makeatletter
\newif\if@restonecol
\makeatother

\usepackage{tikz}
\usepackage{pslatex}
\usepackage[arrow, matrix, curve]{xy}
\usepackage{epsfig}
\usepackage[pdfmark]{thumbpdf}
\usepackage{hyperref}
\usepackage{url}

\DeclareMathOperator*{\argmax}{arg\,max}

\urldef{\mailsa}\path|tagiew@informatik.tu-freiberg.de|
\newcommand{\keywords}[1]{\par\addvspace\baselineskip
\noindent\keywordname\enspace\ignorespaces#1}

\begin{document}
\graphicspath{{.}}

\title{Mining Determinism in Human Strategic Behavior}
\author{Rustam Tagiew}
\titlerunning{Mining Determinism in Human Strategic Behavior}
\authorrunning{R.Tagiew}
\institute{Institute for Computer Science of TU Bergakademie Freiberg, Germany\\
\mailsa}
\maketitle
\begin{abstract}
This work lies in the fusion of experimental economics and data mining. It continues author's previous work on mining behavior rules of human subjects from experimental data, where game-theoretic predictions partially fail to work. Game-theoretic predictions aka equilibria only tend to success with experienced subjects on specific games, what is rarely given. Apart from game theory, contemporary experimental economics offers a number of alternative models. In relevant literature, these models are always biased by psychological and near-psychological theories and are claimed to be proven by the data. This work introduces a data mining approach to the problem without using vast psychological background. Apart from determinism, no other biases are regarded. Two datasets from different human subject experiments are taken for evaluation. The first one is a repeated mixed strategy zero sum game and the second -- repeated ultimatum game. As result, the way of mining deterministic regularities in human strategic behavior is described and evaluated. As future work, the design of a new representation formalism is discussed.

\setcounter{page}{84}
\end{abstract}
\section{Introduction}
\indent Game theory is one of many scientific disciplines predicting outcomes of social, economical and competitive interactions among humans on the granularity level of individual decisions \cite[p.4]{tagiewPhD}. People are assumed to be autonomous and intelligent, and to decide according to their preferences. People can be regarded as rational, if they always make decisions, whose execution has according to their subjective estimation the most preferred consequences \cite{russel,rubin}. The correctness of subjective estimation depends on the level of intelligence. Rationality can justify own decisions and predictions of other people's decisions. If interacting people satisfy the concept of rationality and apply mutually and even recursively this concept, then the interaction is called strategic interaction (SI). Further, game is a notion  for the formal structure of a concrete SI \cite{neumann}. A definition of a game consists of a number of players, their legal actions and players' preferences. The preferences can be replaced by a payoff function under assumed payoff maximization. The payoff function defines each player's outcome depending on his actions, other players' actions and random events in the environment. The game-theoretic solution of a game is a prediction about the behavior of the players aka an equilibrium. The assumption of rationality is the basis for an equilibrium. Deviating from an equilibrium is beyond rationality, because it does not maximize the payoff. Not every game has an equilibrium. However, there is at least one mixed strategies equilibrium (MSE) in finite games \cite{nash}.\\
\indent The notion of game is commonly used for pleasant time spending activities like board games, but can also be extended to all social, economical and competitive interactions among humans. A board game can have the same game structure as a war. Some board games are even developed to train people, like Prussian army war game ''Kriegspiel Chess'' \cite{wargaming} for officers. We like it to train ourselves in order to perform better in games \cite{ggp}. In most cases, common human behavior in games deviates from game-theoretic predictions \cite{poolgame,vspt}. One can say without any doubt that if a human player is trained in a concrete game, he will perform close to equilibrium. But, a chess master does not also play poker perfectly and vice versa. On the other side, a game-theorist can find a way to compute an equilibrium for a game, but it does not make a successful player out of him. There are many games we can play; for most of them, we are not trained. That is why it is more important to investigate our behavior while playing general games than playing a concrete game on expert level. Conducting experiments for gathering data of human game playing is called experimental economics.\\
\indent Although general human preferences are a subject of philosophical discussions \cite{humannat}, game theory assumes that they can be captured as required for modeling rationality. Regarding people as rational agents is disputed at least in psychology, where even a scientifically accessible argumentation exposes the existence of stable and consistent human preferences as a myth \cite{sociokritik}. The problems of human rationality can not be explained by bounded cognitive abilities only. ''British people argue that it is worth spending billions of pounds to improve the safety of the rail system. However, the same people habitually travel by car rather than by train, even though traveling by car is approximately 30 times more dangerous than by train!''\cite[p.527--530]{hrational} Since the last six decades nevertheless, the common scientific standards for econometric experiments are that subjects' preferences over outcomes can be insured by paying differing amounts of money \cite{performmoney}. However, insuring preferences by money is criticized by the term ''Homo Economicus'' as well.\\
\indent The ability of modeling other people's rationality and reasoning as well corresponds with the psychological term ''Theory of Mind'' (ToM) \cite{verbrugge}, which lacks almost only in the cases of autism. For experimental economics, subjects as well as researchers, who both are supposed to be non-autistic people, may fail in modeling of others' minds anyway. In Wason task at least, subjects' reasoning does not match the researchers' one \cite{wason}. Human rationality is not restricted to capability for science-grade logical reasoning -- rational people may use no logic at all \cite{nonlogic}. However, people also mistake seriously in the calculus of probabilities \cite{tversky}. In mixed strategy games, the required sequence of random decisions can not be properly generated by people \cite{randomexp}. Due to bounded cognitive abilities, every ''random'' decision depends on previous ones and is predictable in this way. In ultimatum games \cite[S. 43ff]{vspt}, the economists' misconception of human preferences is revealed -- people's minds value fairness additionally to personal enrichment. Our minds originated from the time, when private property had not been invented and social values like fairness were essential for survival.\\
\indent This work concentrates on human playing of general games and continues author's previous work \cite{tagiewaiia}. It is about the common human deviations from predicted equilibria in games, for which they are not trained or experienced. The two examples introduced in this work are a repeated mixed strategy zero sum game and a repeated ultimatum game from responders' perspective. The only assumption is the existence of deterministic rules in human behavior. Under this assumption, diverse data mining algorithms are evaluated. Apart from mining deterministic regularities, modeling human behavior in general games needs a representation formalism which is not specific to a concrete game. Representing human behavior models in such a formalism would increase their comparability. Therefore, this paper includes a general formalism discussion, where results from the evaluation are involved.\\
\indent The next section summarizes related work on a formalism for human behavior in games. Then, the data mining approach on datasets is presented afterwards. Summary and discussion conclude this paper.\\
\section{Related Work}
\indent A very comprehensive gathering of works in experimental psychology and economics on human behavior in general games can be found in \cite{vspt}. This work inspired research in artificial intelligence \cite{aviNID}, which led to the creation of network of influence diagrams (NID) as a representation formalism. NID is a formalism similar to the possible worlds semantics of Kripke models \cite{kripke} and is a super-set of Bayesian games. The main idea of NID is modeling human reasoning patterns in diverse SIs. Every node of a NID is a multi-agent influence diagram (MAID) representing a model of SI of an agent. MAID is an influence diagram (ID), where every decision node is associated with an agent. ID is a Bayesian network (BN), where one has ordinary nodes, decision nodes and utility nodes. In summary, this approach assumes that human decision making can be modeled using BN -- human reasoning is assumed to have a non-deterministic structure. This formalism is already applied for modeling reciprocity in a repeated ultimatum game called ''Colored Trails'' (CT) \cite{aviNegot}. The result of this work is that models of
adaptation to human behavior based on BNs perform better than standard game theoretical algorithms.\\
\indent Another independent work is an application of a cognitive architecture from psychology to games \cite{actrgames}. A cognitive architecture is a formalism concerned to represent general human reasoning \cite{cogarchitek} in order to compare different models. Today's most popular cognitive architecture is ACT-R (Adaptive Control of Thought — Rational) \cite{actr}. In comparison to NID, ACT-R is used for a number of psychological studies. ACT-R consists of two tiers -- symbolic and sub-symbolic. On the symbolic tier, there are chunks -- facts and ''If-Then''-rules. On the sub-symbolic tier, there are exponential functions, which determine activation levels of chunks, delays in reasoning and priorities between rules. Based on ACT-R, an almost deterministic model for a mixed strategy zero sum game ''Rock Paper Scissors'' (RPS) is designed. The only case, in which the designed model predicts random behavior is the beginning of a game sequence. The model was successfully evaluated as a base for an artificial player, which won against human subjects.\\
\indent Whether deterministic or not, both works follow the same approach. First, they construct a model, which is based on theoretical considerations. Second, they adjust the parameters of this model to the experimental data. This makes the human behavior explainable using the concepts from the model, but needs a priori knowledge to construct the model.
\section{Used Datasets}
\indent The first dataset chosen for our data mining approach has already been mentioned in our previous  work \cite{tagiewaiia}. It is the game RPS played over a computer network. This game is easy to explain and most people do not train to play it on expert level; it is symmetric, zero sum and two player. The study was conducted on threads of $30$ one-shot games. A player had a delay for consideration of $6$ sec for every shot. If he did not react, the last or default gesture was chosen. A thread lasted $30*6\text{ sec } = 3 \text{ min}$. This game has one mixed strategy equilibrium (MSE), which is an equal probability distribution between the three gestures. At least, one can not lose playing this MSE.\\
\indent Ten computer science undergraduates were recruited. They were in average $22,7$ years old and $7$ of them were male. They had to play the thread twice against another test person. Between the two threads, they played other games. In this way, $300$ one-shot games or $600$ single human decisions are gathered. Every person got \EUR{0.02} for a won one-shot game and \EUR{0.01} for a draw. The persons, who played against each other, sat in two separate rooms. One of the players used a cyber-glove and the other one a mouse as input for gestures. The graphical user interface showed the following information - own last and actual choice, opponents last choice, a timer and already gained money. According to statements of the persons, they had no problems to understand the game rules and to choose a gesture timely. All winners and $80$\% of losers attested that they had fun to play the game.\\
\indent The second dataset is the recorded responder behavior from the CT experiment \cite{aviNegot}. This dataset contains $371$ single human decisions of $25$ participating subjects. A positive decision of the responder updates the monetary payoff of both players, while a negative one does not change anything. The payoff update varied between \${1.45} and \${-1.35} for the responder. In $160$ cases, responders update was zero. The equilibrium for the responder is to accept only proposals, which increase his payoff regardless of the proposer's payoff.\\
\section{Methods}
\indent Statistical analysis of the datasets from the previous chapter exposed that the human behavior observed in the experiments can not be explained using only game theory \cite{tagiewPhD}. The shape of equilibrium deviations confirms the one reported in relevant literature \cite{vspt}. The goal is to find a model beyond game theory for the prediction of average deviations. In related work, the creation of a sophisticated model preceded the evaluation on the data.  In this work, the evaluation on the data precedes the creation of a model. Of course, some people would not match into such a model like trained or somehow experienced individuals. Prediction of specific individuals is not addressed in this paper.\\
\indent Machine done prediction without participation in game playing with human subjects should not be confused with prediction algorithms of artificial players. Quite the contrary, artificial players can manipulate the predictability of human subjects by own behavior. For instance, an artificial player, which always throws ''Paper'' in RPS, would success at predicting a human opponent always throwing ''Scissors'' in reaction. Otherwise, if an artificial player maximizes its payoff based on opponent modeling, it would face a change in human behavior and have to handle that. This case is more complex than a spectator prediction model for an ''only-humans'' interaction. This paper restricts on a prediction model without participating.\\
\indent Human behavior can be modeled as either deterministic or non-deterministic. Although human subjects fail at generating truly random sequences as demanded by MSE, non-deterministic models are especially used in case of artificial players in order to handle uncertainties. ''Specifically, people are poor at being random and poor at learning optimal move probabilities because they are instead trying to detect and exploit sequential dependencies. ... After all, even if people do not process game information in the manner suggested by the game theory player model, it may still be the case that across time and across individuals, human game playing can legitimately be viewed as (pseudo) randomly emitting moves according to certain probabilities.'' \cite{actrgames} In the addressed case of spectator prediction models, non-deterministic view can be regarded as too shallow, because deterministic models allow much more exact predictions. Non-deterministic models are only useful in cases, where a proper clarification of uncertainties is either impossible or costly. To remind, deterministic models should not be considered to obligatory have a formal logic shape.\\
\indent The deterministic function $\text{HD}(\text{Game}, \text{History})\rightarrow\text{Decision}$ denotes a human decision. $History$ denotes the previous turns in the game. $Game$ and $History$ are the input and $Decision$ -- the output. Finding a hypothesis, which matches the regularity between input and output without a priori knowledge, is a typical problem called supervised learning \cite{ml}. There is already a big amount of algorithms for supervised learning. Each algorithm has its own hypothesis space ($\text{HS}$). For a Bayesian learner, e.g., the hypothesis space is the set of all possible Bayesian networks. There are many different types of hypothesis spaces - rules, decision trees, Bayesian models, functions and so on. Concrete hypothesis $\text{HD}^{I}$ is a relationship between input and output described by using the formal means of the corresponding hypothesis space.\\
\indent Which hypothesis space is most appropriate to contain valid hypotheses about human behavior? This is a machine learning version of the question about a formalism for human behavior. The most appropriate hypothesis space contains the most correct hypothesis for every concrete example of human behavior. A correct hypothesis does not only perform well on the given data (training set), but it performs also well on new data (test set). Further, it can be assumed that the algorithms which choose a hypothesis perform alike well for all hypothesis spaces. For instance, a decision tree algorithm creates a tree, a neuronal algorithm creates a neuronal network and the distance between the created tree to the best possible tree is the same as the distance between the created neuronal network and the best possible neuronal network. This assumption is a useful simplification of the problem for a preliminary demonstration. Using it, one can consider the algorithm with the best performance on the given data as the algorithm with the most appropriate hypothesis space. The standard method for measurement of performance of a machine learning algorithm or also a classifier is cross validation.\\
\indent As it is already mentioned, a machine learning algorithm has to find hypothesis $\text{HD}^{I}$ which matches best the real human behavior function $\text{HD}$. Human decision making depends mostly on a small part of the history due to bounded resources. This means that one needs a simplification function $\text{S}(History)\rightarrow\text{Pattern}$. Using function $\text{S}$ the function $\text{HD}(X,Y)$ is to be approximated through $\text{HD}^{II}(X,S(Y))$. The problem for finding the most appropriate hypothesis can be formulated in equation \ref{e01}. The function $\text{match}$ in equation \ref{e01} is considered to be implemented through a cross validation run.\\
\begin{equation}\label{e01}
\argmax_{\text{HS}}(\max_{\text{HD}^{II}\in\text{HS}}(\text{match}(HD(X,Y),\text{HD}^{II}(X,S(Y)))))
\end{equation}
\section{Empirical Results}
\indent The first dataset is transformed to a sets of tuples, each one consists of three own previous gestures, three opponent previous gestures and own next gesture. Therefore, every tuple has the length $3+3+1=7$. The simplification function is a window over three last turns. There are $2187$ possible tuples for RPS. The decisions in the first three turns of game are not considered. Therefore, the size of the set results to $540$ tuples. The second dataset is also transformed to a sets of $371$ tuples, where every tuple includes the proposers payoff update, the responders payoff update and the responders boolean reply.\\
\indent Implementations of classifiers provided by WEKA \cite{Weka} are used for the cross validation on the both sets of tuples. For the first dataset, there are currently $45$ classifiers available in the WEKA library, which can handle multi-valued nominal classes. Gestures in RPS are nominal, because there is no order between them. These classifiers belong to different groups - rule-based, decision trees, function approximators, baysian learners, instance-based and miscellaneous. A cross validation of all $45$ classifiers on RPS dataset is performed. For the CT dataset, a cross-validation of $35$ appropriate classifiers is performed. The number of subsets for cross-validation is $10$. Both cross-validation runs are conducted with preserving order of the tuples.\\
\indent Sequential minimal optimization (SMO) \cite{Platt1998} showed $46.48$\% prediction correctness, which is about $1$\% higher than the sophisticated non-deterministic model for RPS of Warglen \cite{warg}. Unfortunately, decreasing and increasing the window size in the function $\text{S}$ for the RPS dataset diminishes the performance. Using the single rule classifier (OneR), one can find out that $43.15$\% of the RPS dataset matches the rule: ''Choose paper after rock, scissors after paper and rock after scissors''. A number of classifiers including SMO achieve $95.42$\% correctness on the CT dataset in cross-validation. One of this algorithms is based on decission tables \cite{decisiontable}. This algorithm finds out that $95.15$\% of the CT dataset conforms the rule: ''If an acceptance does neither change your payoff nor improve the proposers payoff, then refuse!'' This result overperforms clearly the $72$\% reported from the non-deterministic approach of Pfeffer \cite{aviNegot}.\\
\section{Conclusion}
\indent The strategic behavior consists out of the observable actions, whose origins are tried to be understood as generally as possible. Summarizing the results of this work, it can be said that SMO can find the most general deterministic hypothesis about regularities of human behavior in the investigated scenarios. The correctness of such a hypothesis overperforms the numbers reported in related work. The hypothesis space of SMO is one of complex functions and can be used for the design of a game behavior description formalism.
\bibliographystyle{splncs}
\bibliography{human03}
\end{document}